# Structural Transformation of Implanted Diamond Layers during High Temperature Annealing

S. Rubanov[1], B. A. Fairchild[2,3], A. Suvorova[4], P. Olivero[2,5] and S. Prawer[2]

[1] Bio21 Institute, University of Melbourne, Australia
[2] School of Physics, University of Melbourne, Australia
[3] MNRF, RMIT University, Melbourne, Australia
[4] CMCA, the University of Western Australia, Australia
[5] Physics Department, University of Torino, Torino, Italy

## 1. Introduction

The progress in the fabrication of synthetic diamond has resulted in increasing number of its potential applications. Single crystal diamond has also attracted enormous interest as a solid state platform for quantum information processing. Nitrogen-vacancy (N-V) color centers in diamond show remarkable quantum properties such as long coherence times and single spin readout, and can be used as qubits in a quantum computer architecture [1-3]. In order to take advantage of these properties, it is highly desirable to fabricate photonic components in diamond at the micro and even nano-scale level. In a previous work we demonstrated the ability of fabricating three-dimensional structures in diamond at the micro-scale level using lift-out method [4]. MeV ion implantation was used to create a buried damage layer which transformed into a graphite-like layer upon high temperature thermal annealing. The graphitic layer can be selectively etched to form a free-standing membrane into which the desired structures can be sculpted using focused ion beam (FIB) milling. Ion implantation with tens of keV ion energy [5] or multiple energy implantation techniques [6] when combined with FIB milling allows device fabrication in diamond at the micro- and nano-scale. The modeling of the optical properties of the devices fabricated using this method [4-8] are based on the assumption of a sharp diamond-air interface. The real quality of this interface could depend on many factors, for example, on the degree of graphitization of the amorphous damage layers after annealing. Kalish et al. [9] reported complete graphitization of the implanted layer in diamond after a 20 min annealing at 600 °C. However, a recent study [10] using high-resolution electron microscopy (HREM) revealed the presence of a transition area with pockets of crystalline diamond in the graphite matrix near surface, after high temperature vacuum annealing of implanted layer in diamond. Thus, the real quality of the diamond surface after chemical removal of graphitic layer can be far from ideal. In the present work the processes of the ion-beam-induced amorphisation and graphitisation in diamond were studied using cross-sectional conventional and analytical TEM. The graphitisation of the amorphized layers was carried out at annealing temperatures of 550 $^{o}$C and 1400 $^{o}$C.

## 2. Experimental

Synthetic (001) diamond samples produced by Sumitomo Inc. were implanted at room temperature with He$^{+}$ ions to a fluence range of $3 \times 10^{16}$-$10 \times 10^{16}$ cm$^{-2}$. The energy of the He$^{+}$ ions was 0.5 MeV or 2 MeV. The ion implantation with ion energy of 0.5 MeV was performed through a mask, (a copper TEM mesh grid). Thus, the latter sample contains implanted areas separated by unimplanted areas screened during implantation by grid bars. The 0.5 MeV implantation was chosen to create the damage layer in diamond much closer to the specimen surface for TEM imaging of the cross-section of implanted and unimplanted regions at the same time.

The samples were annealed for 1 hour in forming gas (4% hydrogen in argon) atmosphere at 550 $^{o}$C and in vacuum at 1400 $^{o}$C. Cross sectional TEM samples were prepared using the lift-out method [11-12] after ion implantation and after thermal annealing. Prior to TEM specimen preparation, all implanted diamond samples were coated with thin carbon films and 300 nm thick Pt protection layers were deposited using electron-beam deposition facility of the FEI Nova Nanolab dual-beam FIB system in order to mask the TEM cross section from the Ga beam. Cross-sectional TEM samples were prepared in [110] and [100] orientations. Also, a cross-sectional TEM sample was prepared from the sample after annealing at 550 $^{o}$C and chemical etching of graphitic layer in boiling acid (1:1:1 H$_2$SO$_4$/HClO$_4$/HNO$_3$). This TEM lamella contains the bottom interface of the diamond cap layer (i.e. the layer comprised between the surface and the buried heavily damaged layer). Conventional TEM imaging was done using a Tecnai TF 20 electron microscope operated at 200 kV. Energy-filtered TEM (EFTEM) imaging was conducted at 200 kV (JEOL, JEM-2100). Electron energy loss spectroscopy (EELS) was conducted at 300 kV (JEOL 3000F).

## 3. Results and discussions

The interaction of the energetic ions with the diamond substrate initiates a sequence of displacement events that leads to the production of lattice defects (vacancies and interstitials) and, at sufficiently high fluences, to the crystalline-to-amorphous (c-a) transformation of the irradiated volume [4, 6]. The amorphous damage layer after 0.5 MeV $He^+$ ion implantation is clearly visible in the bright-field TEM image in Fig. 1a due to absence of the diffraction contrast or long-range order in this area. The depth of the heavily damaged layer correlates in general with damage profile calculated using the SRIM 2008 Monte Carlo code [13]. With the exception of the area in the vicinity to the right edge, the amorphized damage layer has uniform thickness (200±10) nm. The surface height step (75±4 nm) between unimplanted (masked by TEM grid bar) and implanted regions is evident and is indicated by the white line and the arrow in Fig.1a. The conducted EELS measurements have shown the presence of both $sp^2$ and $sp^3$ bonding in the implanted region. The swelling of the implanted layer has been attributed to the conversion of the diamond's $sp^3$ bonds to graphite-like $sp^2$ bonds with significant decrease in density. The positive step height between unimplanted and implanted regions could be a result of a Poisson ratio effect arising from the biaxial compressive stress in the plane of the implanted layer [14].

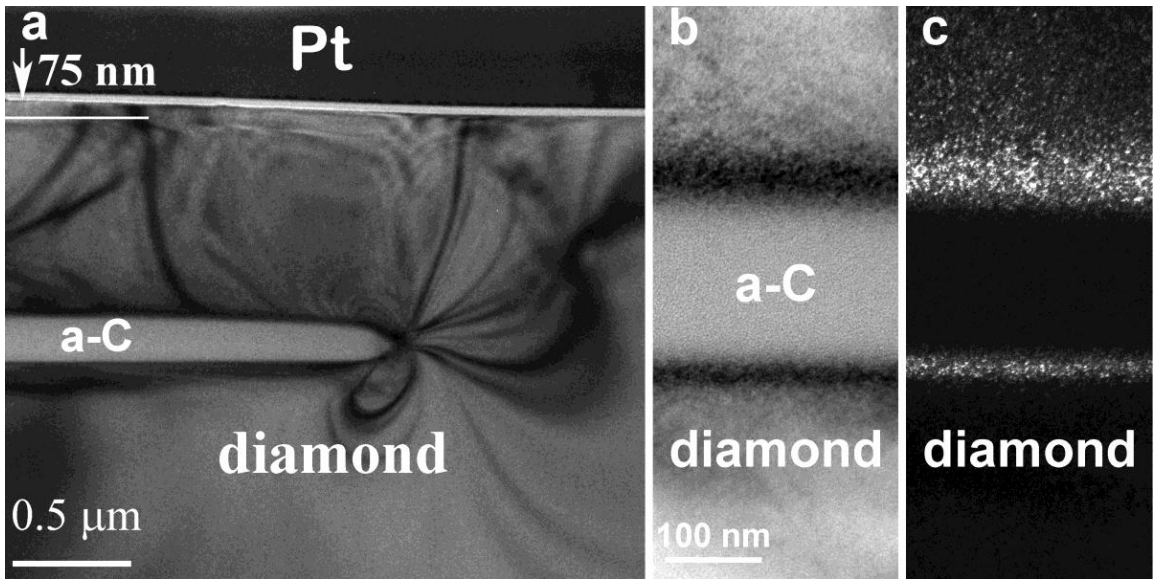

Fig.1. a) Cross-sectional bright-field TEM image of the diamond sample near the edge of 0.5 MeV implantation area; b) bright field image of damage layer in diamond after 2 MeV He$^+$ implantation; c) corresponding weak beam dark field image.

Dark strain contours can be noticed along both amorphous-crystal interfaces in Figs.1a and 1b. These contours represent strain field in the implanted region. Fig. 1b,c shows magnified bright field and **g**-3**g** weak beam dark field (WBDF) images with diffraction vector **g** = [2-20] from an erea implanted with 2 MeV He$^+$. WBDF is a powerful method for imaging crystal defects like dislocations with a high spatial resolution. Only near the core of the lattice defects the strain is large enough to bend the crystal planes back to the Bragg condition. The contrast in the WBDF image is very strong and sharp in the regions where the crystal lattice is distorted and this area is visible as a band of small white spots in Fig. 1c. However, it is worth noting that despite of the large number of the lattice defects diamond remains crystalline in this region. The distorted region is thicker for the leading edge of the implanted area and thinner for tailing edge, which correlates with damage profile as simulated by SRIM. The concentration of point defects here is below the critical value which is known as the amorphisation threshold ($D_c$) but high enough to cause local crystal lattice distortion. Depending on the fabrication technique, amorphous carbon films can have a wide range of diamond-like, sp$^3$ bonded carbon content, ranging from pure sp$^2$ to approximately 80% sp$^3$ (tetrahedral amorphous carbon) [15]. EELS measurements of the carbon K-edge in the amorphous damaged region showed a prominent feature at 285 eV, i.e. the $\pi^*$ peak associated with the presence of sp$^2$ bonding [16]. This, together with swelling, indicates the conversion of the diamond sp$^3$ bonds to sp$^2$ in the amorphous damage area. Using a mass-balance calculation with the values of height step and amorphous layer thickness (Fig. 1), the average density of the amorphous damage layer after ion implantation in diamond was calculated to be ~2.2 g·cm$^{-3}$ (~20% sp$^3$ fraction), a value which is close to the density of graphite (2.09 – 2.23 g·cm$^{-3}$).

Fig. 2a shows the damage layer in diamond implanted with 2 MeV He$^+$ ions after 1 hour annealing at 550 $^o$C. A new phase is visible with brighter contrast in the middle of the damage layer. Selected-area diffraction patterns taken from the area containing this damage layer from the specimens prepared in [100] and [110] orientations are shown in Fig. 2b and 2c. The diameter of the selected area aperture was slightly bigger than the width of the damaged layer covering interfaces with diamond and diffraction spots from the diamond are also present (some are identified in these pictures).

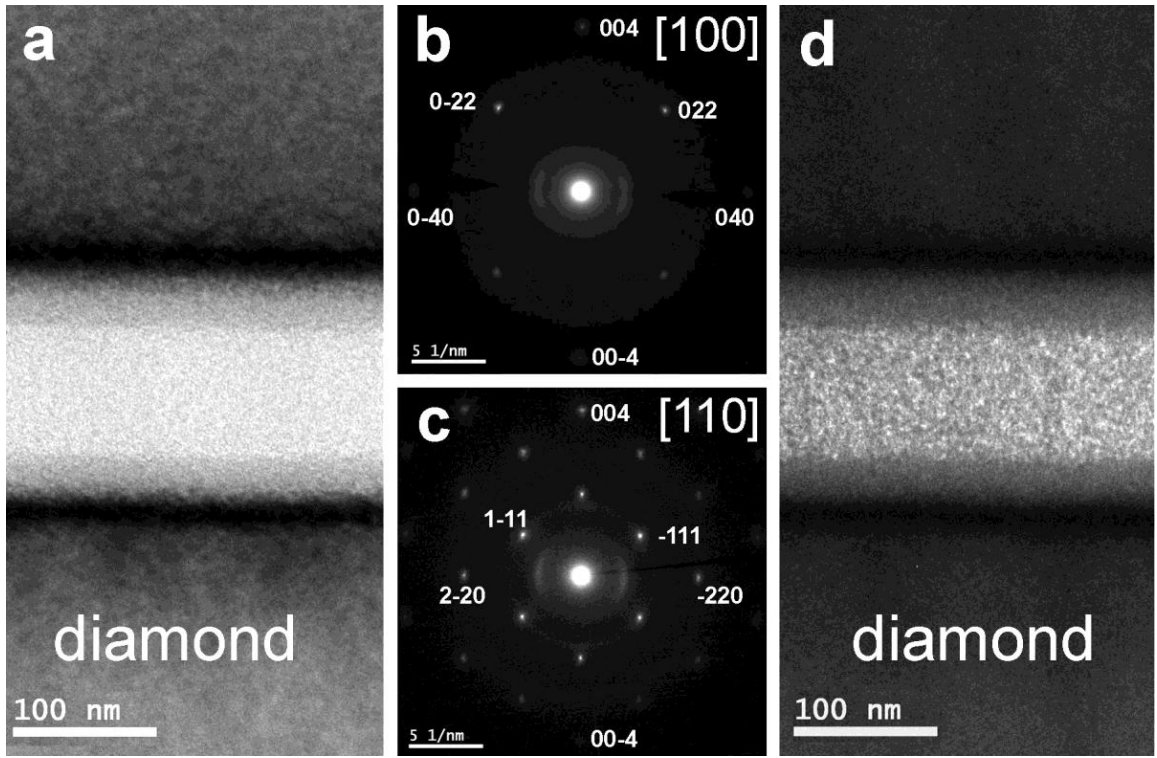

Figure 2 a) [100] cross-sectional bright field image of the damage layer in diamond after 2 MeV He$^+$ implantation and 1 hour annealing at 550 $^o$C; b) selected area diffraction pattern from the damage layer for zone axis [100]; c) same for zone axis [110]; d) a dark field image of the damage layer, **g** = [002] of graphite.

Two arcs in the diffraction pattern are also visible (Figs. 2b and 2c), which correspond to graphite (002) atomic planes. These arcs indicate the presence of the graphite nano-crystals which are oriented semi-randomly with some preferred orientation. Such preferential orientation has graphite (002) planes parallel to diamond (040) atomic planes for diffraction into [100] direction (Fig. 2b) and parallel to diamond (220) atomic planes for diffraction into [110] direction (Fig. 2c). This means that graphite crystals in the damaged layer after annealing are oriented semi-randomly with (002) atomic planes arranged predominantly vertical to the diamond surface. The dark field image of the damage layer with diffraction vector [002] of graphite is shown in Fig. 2d. Graphite nano-crystals are now visible as the bright spots. The average size of the graphite nano-crystals does not exceed 5 nm. These nano-crystals are only visible in the central part of the damage layer. This means that at the given annealing conditions the graphitisation took place only in the central part of the damage layer which after ion implantation has lowest density and highest concentration of $sp^2$ bonds [17].

The damage layer after annealing at 550 °C was also examined using EELS and energy-filtered TEM. Fig. 3a shows EELS carbon K-edge from diamond region, interface region and the central region of the implanted layer shown in Fig. 2. The carbon K-edge obtained from the central and interface region shows a prominent feature at 285 eV, characteristic of transitions to $\pi^*$ states in $sp^2$ bonded carbon and indicating that graphitic or amorphous carbon is present in the layer [16]. The $\pi^*$ peak associated with the presence of $sp^2$ bonding allows the mapping of the variation in intensity in this feature as a function of position within the microstructure, as it has been demonstrated for mapping $sp^2$ bonded carbon by Muller et al. [18].

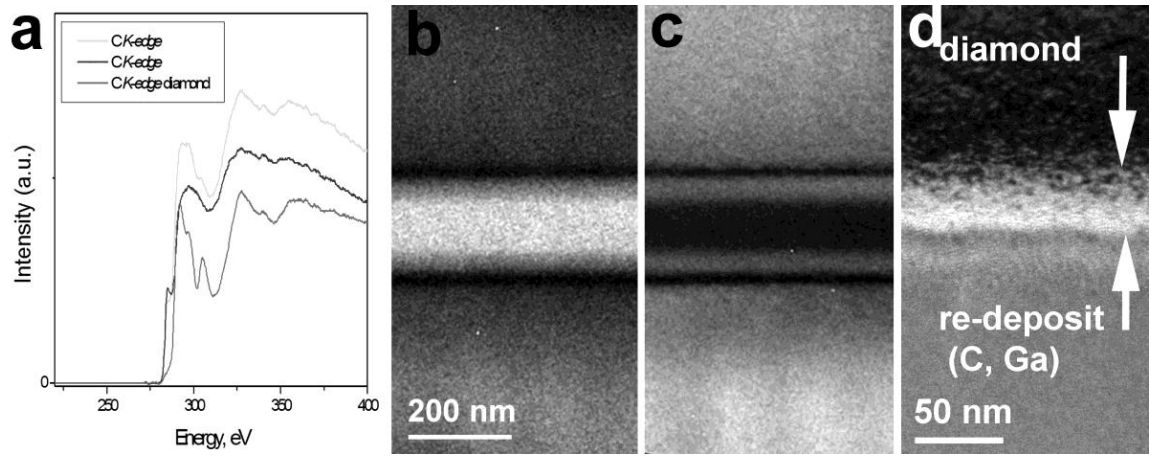

Fig. 3. a) EELS spectrum of carbon K-edge in central region of 2 MeV He$^+$ implanted layer, near interface and in diamond after annealing at 550 °C; b) The chemical state mapping at 285 eV and c) 290 eV; d) interface area after chemical etching.

We used chemical state mapping positioning a narrow energy window (2 eV) over the $sp^2$ and $sp^3$ peaks at 285 eV and 290 eV respectively to record the spatial maps shown in Figs. 3b and 3c. Energy-selected images of the $sp^2$ and $sp^3$ carbon at 285 eV and 290 eV clearly revealed that the implanted layer consists of two different phases of carbon.

The TEM results (Fig. 2) are consistent with the images obtained by the mapping across the implanted region. The central part of the damaged layer after 1 hour annealing at 550 °C is fully converted to the nanocrystalline graphite. **The rest of the damaged layer** remains amorphous with large fraction of sp$^3$-bonded carbon atoms (tetrahedral amorphous carbon). During the fabrication of the thin diamond layer [4, 5] only graphitic phase could be chemically etched away. Figure 3d shows the cross-sectional image of the bottom interface of the diamond cap layer after chemical etching of the graphitic layer. During the TEM sample preparation the interface was covered re-deposited material, which contained C and Ga atoms. The interface has an amorphous structure but due to the presence of the Ga atoms it is haracterised by darker contrast with respect to the remaining damaged layer near the diamond interface indicated by white arrows. Thus, after chemical etching the diamond thin films produced with annealing temperature 550 $^o$C could be covered with a layer of the tetrahedral amorphous carbon. The presence of this layer could significantly affect the optical properties of the devices created. Further annealing at temperature 550 $^o$C for 2 hrs did not increase the degree of graphitization, also it is worth remarking that a detailed study on the annealing time dependence of graphitization goes beyond the scopes of this paper.

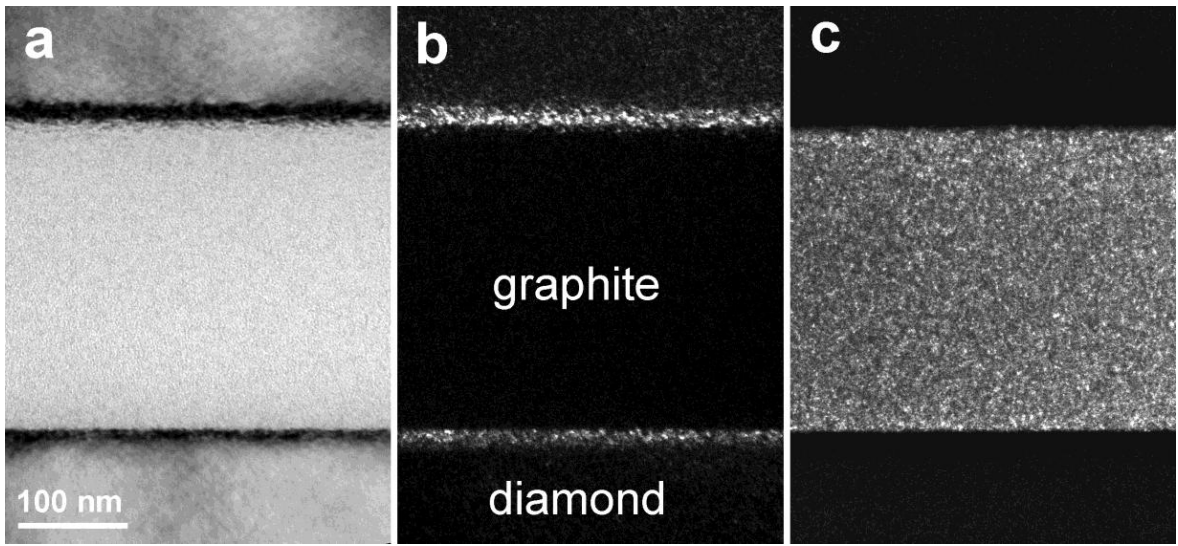

Figure 4. a) Bright field, b) weak beam dark field and c) dark field images of the damaged layer in diamond after 2 MeV He$^+$ implantation and annealing 1 hour at 1400 $^o$C; **g**=[2-20] diamond (a, b); **g**=[002] graphite.

The annealing at higher temperature results in the higher degree of graphitisation of the amorphous damage layer in diamond. Figs. 4 a, 4b and 4c show respectively the bright field, WBDF and dark field images of the damaged layer after annealing at 1400 $^o$C. The bright field image (Fig. 4a) shows an improvement of the diamond/graphite interface: it becomes sharper, especially in the trailing edge. The damaged layer between the two diamond interfaces is uniform in contrast. The WBDF image (Fig. 4b) revealed that the diamond lattice near both interfaces remained distorted. This means that defect complexes responsible for this lattice distortion are very stable even at high temperature annealing. The entire graphitization of the amorphous damage layer is evident in Fig. 4c. The graphitic nano-crystals are uniformly distributed across the damage layer. The average size of these nano-crystals is similar to that of the nano-crystals after 550 $^{o}$C annealing. Graphite particles are visible in implanted layers with *c*-planes predominantly oriented perpendicular to the surface. This correlates with thermodynamic calculations which predict that a biaxial compressive stress will orient the graphitic particles with their *c* axis perpendicular to the stress field [14]. However, the lattice distortion is still present in diamond near top and bottom interfaces, as visible in the WBDF image (Fig. 4b), indicating the absence of solid phase recrystallisation of diamond during high temperature annealing. This also means that there is no reverse conversion of the sp$^2$ to sp$^3$ bonds during atmospheric pressure annealing. On the other hand the absence of any graphite nano-crystals inside the damaged diamond region in Fig. 4c confirmed the suppression of the process of conversion of broken sp$^3$ bonds into sp$^2$ bonds in the crystalline damaged area. It can be also seen in Figs. 4c that the boundaries of the graphitic layer are very sharp. Some isolated diamond nano-clusters that are still present inside the graphitic layer should be washed away during chemical etching step of the fabrication process. Thus, by using high temperature annealing (1400 $^{o}$C) during fabrication process [4, 5], it is possible to create optical device structures with sharp air-diamond interfaces. However the properties of these interfaces could be different to the ideal case due to the presence of the residual damage.

Also, it was shown recently [19-20] that in contrast to VPHT treatment, implanted layers in diamond after high pressure high temperature (HPHT) annealing (1200÷1600 $^{o}$C, 4÷8 GPa) became graphitic with high degree of crystallinity (multi-layer graphene structure). High pressure during HPHT results in change of stress field in implanted region and introduction of third compressive stress component normal to the diamond surface. This results in the epitaxial regrowth of the amorphous damaged layer into single crystal graphite with *c*-planes parallel to the diamond (111) planes and sharp graphite-diamond interfaces.

Moderate temperature (550 $^{o}$C) at high pressure annealing of the implanted diamond samples could possibly lead to similar results with formation of multi-layer graphene sandwiched between layers of tetrahedral amorphous carbon. Thought, the presence of amorphous carbon layers is unwanted in lift-out process [21], the hetero-structures containing three different forms of carbon – diamond, tetrahedral amorphous carbon and graphite could find new applications in the future.

**Conclusions**

High-fluence MeV He$^{+}$ ion implantation in diamond results in the formation of a large number of lattice defects with corresponding distortion of the diamond lattice and formation of the buried amorphous layer. The thermal annealing at moderate temperature (550 $^{o}$C) resulted in a partial graphitization of the implanted volume and in the formation of the nano-crystalline graphitic phase which is sandwiched between layers of tetrahedral amorphous carbon. Selected area diffraction and dark field imaging revealed that the average size of graphite nano-crystals did not exceed 5 nm with predominant orientation of the *c*-planes normal to the sample surface. Annealing at 1400 $^{o}$C results in complete graphitization of the amorphous damage layer. However, a layer of distorted diamond remains near the interface even after high temperature annealing.